\documentclass[%
 reprint,
 nofootinbib,
 amsmath,amssymb,
 aps,
]{revtex4-2}

\usepackage{graphicx}
\usepackage{dcolumn}
\usepackage{bm}
\usepackage{hyperref}

\begin{document}


\title{Gauge symmetry of unimodular gravity in Hamiltonian formalism}

\author{I. Yu. Karataeva}
\email{karin@phys.tsu.ru}
\author{S. L. Lyakhovich}
\email{sll@phys.tsu.ru}
\affiliation{Physics Faculty, Tomsk State University, Tomsk 634050, Russia
}%

\date{\today}

\begin{abstract}
We work out the description of the gauge symmetry of unimodular gravity in the constrained Hamiltonian formalism.
In particular, we demonstrate how the transversality conditions restricting the diffeomorphism parameters emerge from the algebra of the Hamiltonian constraints.  The alternative form is long known as parametrizing the volume preserving diffeomorphisms by unrestricted two-forms instead of the transverse vector fields. This gauge symmetry is reducible. We work out the Hamiltonian description of this form of unimodular gravity (UG) gauge symmetry.
Becchi-Rouet-Stora-Tyutin--Batalin-Fradkin-Vilkovisky (BFV-BRST)
Hamiltonian formalism is constructed for both forms of the UG gauge symmetry. These two BRST complexes have a subtle inequivalence: Their BRST cohomology groups are not isomorphic. In particular, for the first complex, which is related to the restricted gauge parameters, the cosmological constant does not correspond to any
nontrivial BRST cocycle, while for the alternative complex it does. In the wording of physics, this means $\Lambda$ is a fixed parameter defined by the field asymptotics rather than the physical observable  from the standpoint of the first complex. The second formalism views $\Lambda$ as the observable with unrestricted initial data.
\end{abstract}

\maketitle

\section{Introduction}

\noindent Unimodular gravity (UG) is the version of general
relativity (GR) where the metrics are restricted by the
unimodularity condition,
\begin{equation}\label{detg}
\det{g_{\alpha\beta}}=-1 .
\end{equation}
 Given the restriction, the class of
admissible gauge transformations reduces to the volume preserving
diffeomorphisms,
\begin{equation}\label{Tdiff}
\delta_\xi g_{\alpha\beta}=-\nabla_\alpha\xi_\beta - \nabla_\beta\xi_\alpha ,
\qquad \partial_\alpha\xi^\alpha=0 .
\end{equation}
The Lie brackets of the transverse vector fields are
divergence-free, so the volume preserving diffeomorphisms
(\ref{Tdiff}) form a subgroup in the group of general coordinate
transformations. This subgroup is singled out from the entire
diffeomorphism group  in a special way: The partial differential
equation (PDE) is imposed restricting the gauge parameters
$\xi^\alpha$ rather than the subset being explicitly picked out of the
generators for the gauge subgroup. This is an example of the general
phenomenon of unfree gauge symmetry \cite{KAPARULIN2019114735},
where the gauge variation of the action functional vanishes provided
that the gauge parameters are subject to the PDE system. Among the
other examples of unfree gauge symmetry, we can mention the spin-two
Firtz-Pauli model \cite{Alvarez:2006uu,Blas:2007pp} and
some higher spin field theories
\cite{SKVORTSOV2008301,Campoleoni2013,FRANCIA2014248}. The usual general
theory of the systems with unconstrained gauge parameters cannot be
directly applied to the case of unfree gauge symmetry as the PDEs
restricting the gauge parameters, being essential constituents of
the gauge algebra, have to be accounted for. The restrictions
imposed on the gauge parameters result in modifications of the
second Noether theorem and the Faddeev-Popov quantization rules
\cite{KAPARULIN2019114735}. Also the Batalin-Vilkovisky  (BV)
formalism has to be modified \cite{Kaparulin2019} to account for the
distinctions of the unfree gauge symmetry  from the case of
unconstrained gauge parameters. The specifics of unfree gauge
symmetry in the general constrained Hamiltonian formalism is worked
out in the articles
\cite{Abakumova:2019uoo,Abakumova:2020ajc},
including the modification of the
Hamiltonian
Becchi-Rouet-Stora-Tyutin--Batalin-Fradkin-Vilkovisky (BFV-BRST)
formalism.
A common feature for all the unfree gauge symmetries is that they
admit an alternative formulation with unconstrained gauge parameters
while the unrestricted gauge symmetry is reducible. This general
fact is first noticed in \cite{Kaparulin2019}, though for the
specific models the reducible alternatives have been previously
known.  For the UG, the reducible gauge transformations can be
parametrized by the two-form $W$,
\begin{eqnarray}
 \delta_W g_{{\alpha}{\beta}}=
-\frac{1}{2}\,\varepsilon^{\gamma\sigma\mu\nu}
&\big(&
\partial_{\gamma}g_{{\alpha}{\beta}}\partial_{\sigma} W_{\mu\nu}
+g_{{\alpha}{\gamma}}\partial_{\beta}\partial_{\sigma} W_{\mu\nu}
\nonumber
\\
\label{RedTdiff}
&&
\mbox{}
+g_{\gamma\beta}\partial_{{\alpha}}\partial_{\sigma} W_{\mu\nu}\big),
\end{eqnarray}
where $\varepsilon^{\gamma\sigma\mu\nu}$ is the Levi-Civita symbol.
This
parametrization
of the volume preserving diffeomorphism is long known \cite{Dragon1988K,Unruh1989,Henneaux1989T},
and it is studied once and again mostly at
the linearized level (see
\cite{Alvarez2013H,Kugo:2022iob,Kugo:2022dui},
and references therein).

The transformation (\ref{RedTdiff}) follows from (\ref{Tdiff}) by
substitution of the transverse vector $\xi$ as the Hodge dual of the
exact three-form  $\xi=-*dW$. This substitution can be inequivalent to the
original symmetry (\ref{Tdiff}) if the manifold admits the third
group of  De Rham cohomology. Possible consequences of the
inequivalence are discussed in the article \cite{Abakumova2021KL}.
Here, we do not elaborate on this issue.

Gauge symmetry (\ref{RedTdiff}) is reducible as it admits the
sequence of gauge for gauge transformations:
\begin{eqnarray}
\delta_\varphi W_{{\alpha}{\beta}} &=&
\partial_{{\alpha}} \varphi_{{\beta}}
- \partial_{{\beta}} \varphi_{{\alpha}} ,
\nonumber
\\
\label{Gauge-for-Gauge}
\delta_\psi \varphi_{{\alpha}} &=& \partial_{{\alpha}}\psi .
\end{eqnarray}
The higher spin analogs  of  this form of reducible gauge symmetry
can be found in \cite{FRANCIA2014248}.

At the level of general constrained Hamiltonian formalism, the
procedure of constructing the reducible alternative formulation with
unconstrained gauge parameters has been worked out in the recent
article \cite{Abakumova2021KL} for the general unfree gauge
symmetry. Proceeding from the two forms of gauge symmetry, two
different BRST complexes can be associated with the same action
functional. The first one is for the unfree gauge symmetry, and
another one is for the reducible form of the symmetry. These two
complexes are connected, though they are inequivalent, in general,
in the sense that their BRST-cohomology groups are not necessarily
isomorphic.\footnote{The subtle difference between the cohomology
groups of the two complexes is related to the global conserved
quantities, whose initial data are defined on the lower dimensional
subset, not at the Cauchy surface. All the field theories with
unfree gauge symmetry admit conserved quantities of this type
(see in \cite{Abakumova2021KL}). The simplest example of such a quantity
is the cosmological constant of UG. For the first complex, these
quantities turn out to be a coboundary, while for the alternative
one, they are nontrivial cocycles (see in \cite{Abakumova2021KL}).
In this article, we notice this subtle distinction for the UG in the end of Sec. 4, though we do not elaborate on this fact in the present article.}

The Hamiltonian formulation of GR has a long story. In particular,
various ways have been discussed for decades of reproducing
$4d$-diffeomorphism transformations of the Arnowitt-Deser-Misner (ADM)
 variables, including laps and shift
functions, in Hamiltonian formalism. For pedagogical exposition of
this subject, review, and references, see
\cite{Ponomarev2017BObook,Salisbury2016RS,Salisbury2021}.
The recent discussion of the same issue for the Brans-Dicke
theory can be found in \cite{GabrieleGionti2021}.
Also for the UG, the constrained Hamiltonian
formulation
has been known for at least 30 years
\cite{Unruh1989,Henneaux1989T},
and the topic has been extensively studied
since then; for review and bibliography we refer to
\cite{Percacci2018,Gielen2018AP,Alvarez2021A}.
For the UG, unlike the GR,
it is still  not evident how the Hamiltonian constraints can
reproduce the $4d$ gauge transformation (\ref{Tdiff}) of the theory,
including the transversality condition imposed on the gauge
parameters. This puzzle is reported in the reviews (see, e.g., in
\cite{Gielen2018AP}). Also the reducible form of the UG gauge symmetry
(\ref{RedTdiff}), (\ref{Gauge-for-Gauge}) has never been  described
in Hamiltonian formalism. Once the Hamiltonian description is
lacking for the gauge symmetry (\ref{Tdiff}) or (\ref{RedTdiff}),
(\ref{Gauge-for-Gauge}), the Hamiltonian BFV-BRST formalism is still
unknown for UG. In this article, following the general procedure of
Ref. \cite{Abakumova:2019uoo}, we work out the Hamiltonian
description of the volume preserving diffeomorphisms (\ref{Tdiff}).
We also construct the Hamiltonian description for the alternative
form of the gauge symmetry
(\ref{RedTdiff}), (\ref{Gauge-for-Gauge})
following the general recipe of Ref. \cite{Abakumova2021KL}. Given
the Hamiltonian form of the unfree gauge symmetry for the UG, we
construct the Hamiltonian BFV-BRST complex for the UG following the
prescription of the articles
\cite{Abakumova:2019uoo,Abakumova2021KL} for the general case of unfree gauge
symmetry.

The paper is organized as follows. To make the article
self-contained, in the next section we provide the general basics of
describing the unfree gauge symmetry in Hamiltonian formalism. In
Sec. 3, proceeding from the general scheme of Sec. 2, we
provide Hamiltonian description for the volume preserving
diffeomorphisms and also the Hamiltonian analogs for the reducible
gauge symmetry (\ref{RedTdiff}), (\ref{Gauge-for-Gauge}). Section 4
includes the BFV-BRST complexes for both forms of the gauge symmetry
in UG. The last section includes concluding remarks.

\section{Unfree and reducible gauge symmetry in Hamiltonian formalism}\label{Section2}

In this section, we briefly present the general scheme of deriving
the unfree gauge symmetry for Hamiltonian constrained systems for
the simplest case without tertiary and higher level constraints (the
UG falls in this class of systems). For justification of the scheme,
see in \cite{Abakumova:2019uoo}. The  Hamiltonian description of
unfree gauge symmetry for the most general case with the tertiary
and higher level constraints can be found in
\cite{Abakumova:2020ajc}. The method of finding the alternative
Hamiltonian form of the  unfree gauge symmetry with unrestricted
reducible gauge parameters is worked out in the article
\cite{Abakumova2021KL}. It is briefly explained at the end of this
section.

We begin with the action functional of constrained Hamiltonian system
\begin{eqnarray}
S[q(t),p(t), \lambda (t)] &=& \int dt \big( p_i \dot{q}^i
-H_T(q,p,\lambda)\big),
\nonumber
\\
\label{SH}
\qquad H_T(q,p,\lambda)&=& H(q,p)+
\lambda^\alpha\,T_\alpha(q,p),
\end{eqnarray}
$\alpha= 1,\dots ,m$, where the time dependence is made explicit, while dependence on
space points is implicit. Summation over any condensed index
includes integration over space. The action and the Hamiltonian $H$
are supposed to be integrated over the space. All the constraints are assumed irreducible.

Let us assume the following involution relations of the Hamiltonian $H$, the primary constraints $T_\alpha$, and the secondary ones $\tau_a$:
\begin{eqnarray}
\label{irinv1}
\{T_\alpha,T_\beta\} &=&
U{}^\gamma_{\alpha\beta}(\phi)\,T_\gamma\,,
\\ \label{irinv4}
\{T_\alpha,H\} &=&
V{}^\beta_\alpha(\phi)\,T_\beta
\,+\,V{}^a_\alpha(\phi)\,\tau_a\,,
\\ \label{irinv2}
\{\tau_a,H\} &=&
V{}^\alpha_a(\phi)\,T_\alpha
\,+\,V{}^b_a(\phi)\,\tau_b\,,
\\ \label{irinv3}
\{T_\alpha,\tau_a\} &=&
 U{}^\beta_{\alpha a}(\phi)\,T_\beta
\,+\,U{}^b_{\alpha a}(\phi)\,\tau_b\,,
\\ \label{irinv5}
\{\tau_a,\tau_b\} &=&
U{}^\alpha_{ab}(\phi)\,T_\alpha
\,+\,U{}^c_{ab}(\phi)\,\tau_c\,,
\end{eqnarray}
where the uniform notation is introduced for the phase space
variables $\phi=(q,p)$. These involution relations mean that
conservation of the primary constraints $T_\alpha$ results in the
secondary constraints $\tau_a$ while no tertiary ones arise. The
complete set of constraints is of the first class. The subtlety that
distinguishes the   unfree-generated gauge symmetry from the case of
symmetry with unconstrained gauge parameters lies in the structure
of the coefficient $V^a_\alpha (\phi)$ in involution relation
(\ref{irinv4}). This coefficient is supposed to be a nondegenerate
differential operator in the sense that it has at most a finite
dimensional kernel. The inverse does not exist to $V$ in the class
of differential operators, however. The simplest example of such
$V^a_\alpha$ is a partial derivative. The kernel is one-dimensional
(just any constant), while the inverse is not the differential
operator. If the fields vanish at infinity, there is no kernel at
all, while the operator $\partial_i$ still does not admit a local
inverse. In this case, the secondary constraints $\tau_a$ should
vanish on shell for the sake of consistency of equations of motion,
though relations $\tau_a\approx 0$ are \emph{not differential
consequences} of the primary constraints.

To represent the unfree gauge symmetry transformations in an economic way, it is convenient to introduce $\lambda$-dependent structure functions
\begin{eqnarray}
W_1{}^\beta_\alpha(\phi,\lambda)
&=&
V{}^\beta_\alpha\,-\,U{}^\beta_{\gamma\alpha} \lambda^\gamma ,
\nonumber
\\
W_2{}^\alpha_a (\phi,\lambda)
&=&V{}^\alpha_a\,-\,U{}^\alpha_{\beta a}\lambda^\beta ,
\nonumber
\\
\label{W1W2}
\Gamma{}^b_a{}(\phi,\lambda)
&=&
V{}^b_a\,-\,U{}^b_{\alpha a}\lambda^\alpha\, .
\end{eqnarray}
In terms of these structure functions, the involution relations (\ref{irinv1})--(\ref{irinv3}) read
\begin{eqnarray}
\{\,T_\alpha(\phi), H_T(\phi,\lambda)\}
&=&
W_1{}^\beta_\alpha(\phi,\lambda)\,T_\beta(\phi)
\nonumber
\\
\label{Inv1}
&&\mbox{}+V{}^a_\alpha(\phi)\,\tau_a(\phi) ,
\end{eqnarray}
\begin{eqnarray}
\{\,\tau_a(\phi), H_T(\phi,\lambda)\}
&=&
W_2{}^\alpha_a(\phi,\lambda)\,T_\alpha(\phi)
\nonumber
\\
\label{Inv2}
&&\mbox{}+\Gamma{}^b_a(\phi,\lambda)\,\tau_b(\phi) .
\end{eqnarray}
In this form, the structure is evident of the Dirac-Bergmann
algorithm for the system: Conservation of the primary constraints
(\ref{Inv1}) leads to the secondary constraints $\tau_a$ as the
coefficient $V^a_\alpha$ is nondegenerate even though it is not
invertible in the class of differential operators. Conservation of
the secondary constraints (\ref{Inv2}) does not lead to the tertiary
ones nor does it define any Lagrange multiplier.

As the consequence of involution relations (\ref{Inv1}) and (\ref{Inv2}),
the Hamiltonian action (\ref{SH}) is invariant \cite{Abakumova:2019uoo} under the gauge transformations of phase space variables
$\phi=(q,p)$ and Lagrange multipliers $\lambda^\alpha$,
\begin{equation}\label{Hunfree1}
\delta_\epsilon \phi\,=\,
\{\phi,T_\alpha \}\,\epsilon^\alpha\,+\,\{\phi,\tau_a\}\,\epsilon{}^a,
\end{equation}
\begin{equation}
\label{Hunfree2}
\delta_\epsilon \lambda^\alpha\,=\,
\dot{\epsilon}^\alpha +
W_1{}^\alpha_\beta (\phi,\lambda)\,\epsilon^\beta
+W_2{}^\alpha_a(\phi,\lambda)\,\epsilon{}^a,
\end{equation}
provided that the gauge parameters $\epsilon^\alpha$ and $\epsilon^a$ are subject to the differential equations,
\begin{equation}
\label{HGPE}
\dot{\epsilon}{}^a
\,+\,\Gamma{}^a_b(\phi,\lambda)\,\epsilon{}^b\,
\,+\,V{}^a_\alpha(\phi)\,\epsilon^\alpha
\,=\,0 \,.
\end{equation}
If the structure coefficient $V^a_\alpha$ admitted the inverse,
being a differential operator, the gauge parameters
$\epsilon^\alpha$ could be expressed from (\ref{HGPE}) as
combinations of gauge parameters $\epsilon^a$ and their derivatives.
In this case, we would have the irreducible gauge symmetry with
unconstrained gauge parameters $\epsilon^a$. This symmetry would
involve the second order time derivatives of the gauge parameters.
Once the nondegenerate structure coefficient $V^a_\alpha$ does not
admit any local inverse, the gauge symmetry (\ref{Hunfree1}),
(\ref{Hunfree2}), (\ref{HGPE}) is unfree indeed. To demonstrate that
the above transformation  is a gauge symmetry only under the
restrictions (\ref{HGPE}) imposed on the gauge parameters, let us
compute the gauge variation (\ref{Hunfree1}) and (\ref{Hunfree2}) of
the action (\ref{SH}),
\begin{eqnarray}
\delta_\epsilon S \equiv \int dt \Big(
&&
\big(
\dot{\epsilon}{}^a
\,+\,\Gamma{}^a_b(\phi,\lambda)\,\epsilon{}^b\,
\,+\,V{}^a_\alpha(\phi)\,\epsilon^\alpha
\big)\, \tau_a
\nonumber
\\
\label{direct-var}
&&\mbox{}
-\frac{d}{dt}\big( T_\alpha \epsilon^\alpha + \tau_a \epsilon^a
\big) \Big) .
\end{eqnarray}
Once the secondary constraints $\tau_a$ are assumed to be
irreducible, the variation integrand can reduce to the total
derivative only under the condition that the coefficients vanish at
$\tau_a$. This leads one to impose restrictions (\ref{HGPE}) on
the gauge parameters.

In the next section, the general relations
(\ref{Hunfree1})--(\ref{HGPE})
are specified for the UG reproducing the volume-preserving diffeomorphism in the Hamiltonian setup.

Any Hamiltonian action with unfree gauge symmetry admits the
alternative reducible form of the gauge symmetry with unrestricted
gauge parameters \cite{Abakumova2021KL}. This reducible symmetry
involves the higher order time derivatives of the gauge parameters.
To arrive to this form of gauge symmetry, we introduce an
alternative (overcomplete) set of the secondary constraints
$\widetilde{\tau}_\alpha$ that absorbs the structure coefficients
$V^a_\alpha$,
\begin{equation}\label{tildetau}
\widetilde{\tau}_\alpha \equiv V{}^a_\alpha(\phi)\,\tau_a(\phi) .
\end{equation}
In general, the constraints above are reducible unlike the independent secondary constraints $\tau_a$ (\ref{irinv4}).
In this article, we assume the simplest possible form of the reducibility conditions
\begin{equation}\label{Z}
Z_1{}_A^\alpha\,\widetilde{\tau}_\alpha=0\,,
\qquad
Z_2{}^A_{A_1} Z_1{}_A^\alpha=0\,,
\end{equation}
with  field-independent null vectors $Z_1{}_A^\alpha$ and
$Z_2{}^A_{A_1}$. The general  case with off-shell nontrivial
contributions to the reducibility relations can be found in the
paper \cite{Abakumova2021KL}. Making use of the reducible generating
set of the secondary constraints, the involution relations
(\ref{irinv1})--(\ref{irinv5}) are reorganized as follows:
\begin{eqnarray}
\{T_\alpha,T_\beta\} &=&
U{}^\gamma_{\alpha\beta}(\phi)\,T_\gamma\,,\\
\{T_\alpha,H\} &=&
V{}^\beta_\alpha(\phi)\,T_\beta
\,+\,\widetilde{\tau}_\alpha\,,\\
\{\widetilde{\tau}_\alpha,H\} &=&
\widetilde{V}{}^\beta_\alpha(\phi)\,T_\beta
\,+\,\widetilde{V}{\,}'{}^\beta_\alpha(\phi)\,\widetilde{\tau}_\beta\,,\\
\{T_\alpha,\widetilde{\tau}_\beta\} &=&
\widetilde{\mathcal{U}}{}^\gamma_{\alpha \beta}(\phi)\,T_\gamma\,
\,+\,\widetilde{\mathcal{U}}{\,}'{}^\gamma_{\alpha \beta}(\phi)\,
\widetilde{\tau}_\gamma\,,\\
\{\widetilde{\tau}_\alpha,\widetilde{\tau}_\beta\} &=&
\widetilde{\mathfrak{U}}{}^\gamma_{\alpha\beta}(\phi)\,T_\gamma
\,+\, \widetilde{\mathfrak{U}}{\,}'{}^\gamma_{\alpha\beta}(\phi)\,
\widetilde{\tau}_\gamma\,.
\end{eqnarray}
We introduce $\lambda$-dependent structure functions, being analogs of
(\ref{W1W2}) for the reducible set of constraints:
\begin{equation}\label{W2G2red}
\widetilde{W}_2{}^\beta_\alpha=
\widetilde{V}{}^\beta_\alpha
\,-\,\widetilde{\mathcal{U}}{}^\beta_{\gamma \alpha}\,\lambda^\gamma ,
\qquad
\widetilde{\Gamma}{}^\beta_\alpha=
\widetilde{V}{\,}'{}^\beta_\alpha
\,-\,\widetilde{\mathcal{U}}{\,}'{}^\beta_{\gamma \alpha}\,\lambda^\gamma .
\end{equation}
In terms of these structure functions, the involution relations read
\begin{equation}\label{HTtildetau}
\{\,T_\alpha(\phi), H_T(\phi,\lambda)\}
\,=\,{W}_1{}^\beta_\alpha(\phi,\lambda)\, T_\beta(\phi)\,
\,+\, \widetilde{\tau}_\alpha(\phi) ,
\end{equation}
\begin{eqnarray}
\{\,\widetilde{\tau}_\alpha(\phi), H_T(\phi,\lambda)\}
&=&\widetilde{W}_2{}^\beta_\alpha(\phi,\lambda)\,T_\beta(\phi)
\nonumber
\\
\label{Htildetau2}
&&\mbox{}+\widetilde{\Gamma}{}^\beta_\alpha(\phi,\lambda)\,
\widetilde{\tau}_\beta(\phi) .
\end{eqnarray}
The key distinction of relations (\ref{HTtildetau}) from (\ref{Inv1}) is that reducible secondary constraints $\widetilde{\tau}_\alpha$ \emph{are the differential consequences}  of the primary constraints while the irreducible ones $\tau_a$ are not.
It is the distinction that leads to the reducible gauge symmetry of the action (\ref{SH}) generated by $T_\alpha$ and  $\widetilde{\tau}_\alpha$   with the second time derivatives of unrestricted gauge parameters \cite{Abakumova2021KL}. These
gauge transformations read
\begin{eqnarray}
\delta_\varepsilon \phi
&=&
\{\phi , T_\alpha  \}
\left(
\dot{\varepsilon}{}^\alpha
\,+\,\widetilde{\Gamma}{}^\alpha_\beta \, \varepsilon^\beta
-Z_1{}_A^\alpha\,\varepsilon^A
\right)
\nonumber
\\
\label{RedGS1}
&&\mbox{}
-\{\phi, \widetilde{\tau}_\alpha \}\, \varepsilon^\alpha ,
\end{eqnarray}
\begin{eqnarray}
\delta_\varepsilon \lambda^\alpha
&=&
\left(
\delta^\alpha_\beta\frac{d}{dt}
\,+\, \,{W}_1{}^\alpha_\beta
\right)
\left( \dot{\varepsilon}{}^\beta
\,+\, \,\widetilde{\Gamma}{}^\beta_\gamma\,\varepsilon^\gamma
-Z_1{}_A^\beta\,\varepsilon^A
\right)
\nonumber
\\
\label{RedGS2}
&&\mbox{}
- \widetilde{W}_2{}^\alpha_\beta
\,\varepsilon^\beta  .
\end{eqnarray}
Given the involution relations of reducible constraints
(\ref{Z}), (\ref{HTtildetau}), and (\ref{Htildetau2}),
the above transformations leave the Hamiltonian action (\ref{SH}) invariant modulo a total derivative
while no restrictions are imposed on the gauge parameters
$\varepsilon^\alpha$ and $\varepsilon^A$.

Because of
the reducibility of secondary constraints (\ref{Z}), gauge
transformations  (\ref{RedGS1}) and (\ref{RedGS2}) enjoy a gauge
symmetry of their own. For the simplest case\footnote{For the case
of general reducibility, with off-shell nontrivial contributions,
see \cite{Abakumova2021KL}.} of the constant null vectors, the gauge
transformations of the original gauge parameters read
\begin{equation}\label{SofS1}
\delta_\omega \varepsilon^\alpha = Z_1{}_A^\alpha\,\omega^A ,
\qquad
\delta_\omega \varepsilon^A = \dot{\omega}^A-Z_2{}^A_{A_1}\,{\omega}{}^{A_1}.
\end{equation}
Because of (\ref{Z}), these symmetries are further reducible:
\begin{equation}\label{SofS2}
\delta_\eta \omega^A =
Z_2{}^A_{A_1} \,\eta^{A_1},
\qquad
\delta_\eta \omega^{A_1} = \dot{\eta}^{A_1}.
\end{equation}
In the next section, proceeding from general relations
(\ref{RedGS1})--(\ref{SofS2})
we find the Hamiltonian form of the reducible gauge transformations (\ref{RedTdiff})
and (\ref{Gauge-for-Gauge}) for the UG.

\section{Unimodular Gravity}\label{Section3}
In this section, following the general scheme of the previous one, we construct the transverse diffeomorphism transformations in the Hamiltonian formalism of the UG. We also find the Hamiltonian form of the reducible description for the volume-preserving diffeomorphisms (\ref{RedTdiff}) and (\ref{Gauge-for-Gauge}).

In this section, we use the ADM variables
\begin{eqnarray}
 g_{{\alpha}{\beta}}
 &=&
\begin{pmatrix}
    N^2  + N_k N^k  \ \ & N_j \\
    N_i\ \  & \buildrel\ast\over{g}_{ij} \\
\end{pmatrix},
\nonumber
\\
\label{ADM}
g^{{\alpha}{\beta}}
&=&
\begin{pmatrix}
    N^{-2} \ \  & - N^j N^{-2} \\
   - N^i N^{-2}\ \  & \buildrel\ast\over{g}{\!}^{ij}
   + N^i N^j N^{-2}\\
\end{pmatrix},
\end{eqnarray}
\\
where Latin indices $i,j,k,\dots$ run the values $1,2,3$ and
$N^i=\, \buildrel\ast\over{g}{\!}^{ij}N_j$.
In these variables, the unimodularity condition $\det
g_{{\alpha}{\beta}}=-1$ reads
\begin{equation}\label{Lapse}
N^2 = - \frac{1}{\buildrel\ast\over{g}}\,,
\qquad
\buildrel\ast\over{g} \, = \det \buildrel\ast\over{g}_{ij}.
\end{equation}
This allows one to exclude from the set of variables the laps
function $N$ replacing it\footnote{Recently the modifications of the
UG have been introduced that suggest to replace formula
(\ref{Lapse}) by a more general relation between the lapse and the
$3d$ metrics \cite{BARVINSKY201759,Barvinsky-Kolganov}. The
techniques we propose for describing the gauge symmetry of the
Hamiltonian formalism of UG are not particularly sensitive to the
way of excluding $N$.} by $(-\buildrel\ast\over{g})^{-1/2}$.

In terms of the ADM variables, with $N$ excluded according to (\ref{Lapse}), the Einstein-Hilbert action\footnote{
We use the following definitions for the $3d$ Riemann tensor, Ricci tensor, and scalar curvature:
$ \buildrel\ast\over{R}{\!}^i{}_{j kl}
=\partial_k{\!\!}\buildrel\ast\over{\Gamma}{\!}^i_{lj}
-\partial_l{\!\!}\buildrel\ast\over{\Gamma}{\!}^i_{kj}
+\buildrel\ast\over{\Gamma}{\!}^i_{ks}{\!}
\buildrel\ast\over{\Gamma}{\!}^{s}_{lj}
\\
-\buildrel\ast\over{\Gamma}{\!}^i_{ls}{\!}
\buildrel\ast\over{\Gamma}{\!}^{s}_{kj}
$,
$\buildrel\ast\over{R}_{ij}
=\buildrel\ast\over{R}{\!}^s{}_{i sj}
$, and
$\buildrel\ast\over{R}=
\buildrel\ast\over{g}{\!}^{ij}{\!}\buildrel\ast\over{R}_{ij}.$
} of UG is brought to the Hamiltonian form (\ref{SH}):
\begin{equation}\label{SH-LUG}
\displaystyle S=\int d^4x\,
\big(\Pi^{ij}\partial_0\!\buildrel\ast\over{g}{\!}_{ij}
-\mathcal{H}_T\big), \quad \mathcal{H}_T=\mathcal{H}_0+N^i\,T_i\,,
\end{equation}
where the Hamiltonian reads
\begin{equation}\label{HUG}
\mathcal{H}_0  =- \frac{1}{\buildrel\ast\over{g}}\,
\mathcal{G}_{ij\, kl}
\Pi^{ij}\Pi^{kl}
+\buildrel\ast\over{R}.
\end{equation}
Here, the usual definition is adopted for the De Witt metrics
\begin{equation}
\mathcal{G}_{ij\,kl} =
\frac{1}{2}\left(
\buildrel\ast\over{g}{\!}_{i k}\!
\buildrel\ast\over{g}{\!}_{j l} +
\buildrel\ast\over{g}{\!}_{i l}\!
\buildrel\ast\over{g}{\!}_{j k}
\right)
-\frac{1}{2}\, \buildrel\ast\over{g}{\!}_{ij}\!
\buildrel\ast\over{g}{\!}_{kl} .
\end{equation}
The inverse reads
\begin{eqnarray}
&&\mathcal{G}^{ij\, kl}
=
\frac{1}{2}\, \left(
\buildrel\ast\over{g}{\!}^{ik}\!
\buildrel\ast\over{g}{\!}^{jl}+
\buildrel\ast\over{g}{\!}^{il}\!
\buildrel\ast\over{g}{\!}^{jk}
\right)
 -
\buildrel\ast\over{g}{\!}^{ij}\!
\buildrel\ast\over{g}{\!}^{kl},
\\
\nonumber
&&\mathcal{G}^{ij\,kl}\mathcal{G}_{kl\,sm}
=
\frac{1}{2}(\delta^i_s \delta^j_m +\delta^i_m \delta^j_s).
\end{eqnarray}
The action includes the primary constraints
\begin{equation}\label{TalphaUG}
T_i = - 2 \buildrel\ast\over{g}_{ij}
 \left( \partial_k \Pi^{k  j}
 + \buildrel\ast\over{\Gamma}{\!}^j_{kl}\Pi^{kl} \right),
\end{equation}
with the shift functions $N^i$ serving as the Lagrange multipliers.
The Hamiltonian is a scalar, while the constraints (\ref{TalphaUG})
are covariant $3d$ vector densities.

The requirement of stability \noindent (\ref{Inv1}) of the primary constraints (\ref{TalphaUG}) reads
\begin{eqnarray}
\displaystyle
\left\{T_i\ ,\ \mbox{$\int$} d^3x \,\mathcal{H}_T\right\}
&=&
T_j \,  \partial_i N^j + \partial_j(T_i N^j)
\nonumber
\\
&&\mbox{}+\partial_i \left(\, \frac{1}{\buildrel\ast\over{g}}\,
\mathcal{G}_{sm\, kl}
\Pi^{sm}\Pi^{kl}-\buildrel\ast\over{R} \right)
\nonumber
\\
\label{UGTalpha}
&\approx & \mbox{} 0 .
\end{eqnarray}
The role of the structure coefficient $V^a_\alpha$ [cf.
(\ref{irinv4}) and (\ref{Inv1})] is played by the operator of the partial
derivative. This operator does not admit the local inverse, while
the one-dimensional kernel of $\partial_i$ is formed by constants.
So, we arrive at the single secondary constraint $\tau$ which is
defined by the stability condition (\ref{UGTalpha})  modulo
arbitrary additive constant,
\begin{equation}\label{UGtau}
 \tau = - \frac{1}{\buildrel\ast\over{g}}\,
\mathcal{G}_{ij\, kl}
\Pi^{ij}\Pi^{kl}+ \buildrel\ast\over{R} -\,\Lambda\approx 0\, ,
\quad\Lambda=\mbox{const}.
\end{equation}
The above constraint is the Hamiltonian counterpart of the relation $R=\Lambda$ being the well-known consequence of the UG field equations in Lagrangian formalism
(see, e.g., \cite{Percacci2018}).
The ``integration" constant $\Lambda$ is defined by  the value of the
metric and its derivatives at any single point of space, or by the
asymptotics rather than by initial data at the entire Cauchy
surface. For asymptotically flat space, $\Lambda=0$, for example.
Existence of the  conserved quantities of this type is a common
feature for all the systems with unfree gauge symmetry (see in
\cite{Abakumova2021KL}). The secondary constraint $\tau$ is $3d$
scalar, which differs from the Hamiltonian (\ref{HUG}) by the ``integration"
constant $\Lambda$.  In this sense, the constraint (\ref{UGtau})
means the off-shell conservation of  energy density in the UG, given
the metrics at any single point of the spacetime. This is in line
with the recent discussion of the meaning of the Hamiltonian in the
UG \cite{Alvarez2021A}.

Stability condition for the secondary constraint $\tau$ does not
lead to a tertiary constraint nor does it define any Lagrange
multiplier [cf. (\ref{Inv2})]
\begin{equation}\label{tauUG}
\left\{\tau\ ,\ \mbox{$\int$} d^3x \, \mathcal{H}_T \right\}
=-\partial_i\left((-\buildrel\ast\over{g})^{-1}\!
\buildrel\ast\over{g}{\!}^{ij} T_j  \right) + N^i  \partial_i \tau .
\end{equation}

The involution relations of all the constraints can be
conveniently represented in terms of functionals being contractions
with arbitrary test functions
\begin{eqnarray}
T(\zeta)&=&\int  d^3x \, T_i(x)\,\zeta^i(x) ,
\nonumber
\\
\label{T-funct-unfree}
\tau (\rho')&=&\int  d^3x \, \tau(x)\,\rho'(x) ,
\end{eqnarray}
where $\xi^i(x)$ and $\zeta^i(x)$ are test vector fields and
$\rho'(x)$ and $\sigma'(x)$ are the test functions being arbitrary
scalar densities with weights 1 (letters with strokes).
For these functionals, the
involution relations (equal time P.B.) read
\begin{eqnarray}\label{T-T-UG}
\{ T(\xi)\ ,\ T(\zeta) \}&=&T\big([\xi , \zeta]\big),
\\
\label{T-tau-UG}
\{T(\xi)\ ,\ \tau(\rho') \}
&=&  \tau\big([\xi , \rho']\big),
\\
\label{tau-tau-UG}
\{\tau (\rho')\ ,\ \tau (\sigma') \}&=&
 T\big([\rho',\sigma']\big),
\end{eqnarray}
where
\begin{eqnarray}
&&[\xi , \zeta]^i
=\xi^j\,\partial_j\zeta^i -\partial_j \xi^i\, \zeta^j ,\\
&&[\xi , \rho'] = \partial_i (\xi^i \rho') ,
\quad
[\rho', \xi  ] =- \partial_i (\xi^i \rho') ,\\
&&[\rho',\sigma']^i = \frac{1}{\buildrel\ast\over{g}}\,
\buildrel\ast\over{g}{\!}^{ij}\, (\rho'\, \partial_j  \sigma' -
\partial_j \rho'\, \sigma'  ) .
\end{eqnarray}
Here $[\xi , \zeta]^i$ is a vector, $[\xi , \rho'], [\rho', \xi ] $
--- scalar density with weight 1, $[\rho',\sigma']^i$ --- vector.
The involution relations of the constraints and Hamiltonian read
\begin{eqnarray}\label{49}
\left\{ T_i\ ,\ \mbox{$\int$} d^3x\, \mathcal{H}_0\right\} &=&
-\, \partial_i\tau ,\\
\label{50}
\left\{\tau\ ,\ \mbox{$\int$} d^3x \,\mathcal{H}_0\right\} &=& -\,\partial_i\left((-\buildrel\ast\over{g})^{-1}\!
\buildrel\ast\over{g}{\!}^{ij} T_j  \right) .
\end{eqnarray}

Relations (\ref{UGTalpha}) and (\ref{tauUG}) present the explicit form
of general involutional relations (\ref{Inv1}) and (\ref{Inv2}) in the
case of UG. This allows one to identify the unfree gauge symmetry of
the Hamiltonian UG action (\ref{SH-LUG}) making use of the general
recipe (\ref{Hunfree1})--(\ref{HGPE}). The gauge variations
(\ref{Hunfree1}) and (\ref{Hunfree2})  for the case of UG read
\begin{equation}\label{Hunfree1g-UG}
\delta_\epsilon{\!} \buildrel\ast\over{g}_{ij} =
\epsilon^k \partial_k{\!}\buildrel\ast\over{g}_{ij}
+\buildrel\ast\over{g}_{ik}{\!}\partial_j \epsilon^k
+\buildrel\ast\over{g}_{jk}{\!} \partial_i \epsilon^k
-\frac{2}{\buildrel\ast\over{g}}\, \mathcal{G}_{ij\, kl}
 \Pi^{kl} \epsilon\,,
\end{equation}
\begin{eqnarray}\nonumber
\delta_\epsilon \Pi^{ij}
&=&
\partial_k (\Pi^{ij} \epsilon^k)
-\Pi^{ik}\partial_k \epsilon^j
-\Pi^{jk} \partial_k \epsilon^i
\nonumber
\\
&&\mbox{}+\bigg(\,
\frac{2}{\buildrel\ast\over{g}}
 \buildrel\ast\over{g}{\!}_{kl} \Pi^{ik} \Pi^{jl}
-\frac{1}{\buildrel\ast\over{g}}  \Pi \Pi^{ij}
\nonumber
\\
&&\mbox{}\phantom{@@@}
-\frac{1}{\buildrel\ast\over{g}}\buildrel\ast\over{g}{\!}^{ij}
\mathcal{G}_{kl\,sm}\Pi^{kl}\Pi^{sm}
+\buildrel\ast\over{R}{\!}^{ij} \bigg)  \epsilon\, 
\nonumber
\\
\label{Hunfree1p-UG}
&&\mbox{}-\, \mathcal{G}^{ij\, kl}
\sqrt{-\buildrel\ast\over{g}}
\left(\partial_k\partial_l -
\buildrel\ast\over{\Gamma}{\!}^s_{kl}\partial_s
\right) \frac{\epsilon}{\sqrt{-\buildrel\ast\over{g}}}\,,
\end{eqnarray}
\begin{equation}\label{Hunfree2-UG}
\delta_\epsilon N^i=
\dot{\epsilon}{}^i
+ \epsilon^j\partial_j N^i
-N^j \partial_j \epsilon^i
-\frac{1}{\buildrel\ast\over{g}}
\buildrel\ast\over{g}{\!}^{ij}
\partial_j \epsilon\,,
\end{equation}
where \ $\Pi = \,\buildrel\ast\over{g}_{ij}\!\Pi^{ij}$. The general equations (\ref{HGPE}) constraining the gauge parameters are specialized for UG as
\begin{equation}\label{HGPE-UG}
\dot{\epsilon}+\partial_i(\epsilon^i-N^i\epsilon ) =0\,.
\end{equation}
Relations (\ref{Hunfree1g-UG})--(\ref{HGPE-UG}), being deduced for the case of UG by the general procedure of the previous section,  should represent the gauge symmetry of the Hamiltonian action (\ref{SH-LUG}). Let us verify this fact by directly varying the action given the variations of variables (\ref{Hunfree1g-UG}), (\ref{Hunfree1p-UG}), and (\ref{Hunfree2-UG}),
\begin{eqnarray}
\delta_\epsilon\displaystyle S
&\equiv &
\int d^4x\,
\bigg(
\delta_\epsilon\Pi^{ij}\partial_0{\!}\buildrel\ast\over{g}{\!}_{ij}-
\partial_0\Pi^{ij}\delta_\epsilon{\!}\buildrel\ast\over{g}{\!}_{ij}
-\delta_\epsilon \mathcal{H}_T
\bigg) 
\nonumber
\\
&\equiv & \int d^4x \Big(\big(
\dot{\epsilon}
\,+\,\partial_i (\epsilon^i - N^i   \epsilon )
\big) \tau
\nonumber
\\
&&\phantom{@@@@}
-\partial_0\big( T_i\epsilon^i + \tau \epsilon
\big) \Big).
\end{eqnarray}
As one can see, the gauge variation would  vanish off-shell modulo
total time derivative provided that gauge parameters
$\epsilon$ and $\epsilon^i$ are subject to the Eq.  (\ref{HGPE-UG}).

Notice that the $4d$ diffeomorphism transformation parameters $\xi^\mu$  (\ref{Tdiff}) are connected with their $(1+3)$-split Hamiltonian counterparts
$\epsilon$ and $\epsilon^i$ involved in the transformations (\ref{Hunfree1g-UG})--(\ref{HGPE-UG})  by relations
\begin{equation}\label{xi-epsislon}
  \xi^0= -\epsilon\,,\quad \xi^i=-(\epsilon^i-N^i\epsilon ) \, .
\end{equation}
This locally invertible change of gauge parameters brings the
Eq. (\ref{HGPE-UG}) to the transversality condition
$\partial_\mu\xi^\mu=~0$. Also notice that Hamiltonian gauge
transformations (\ref{Hunfree1g-UG}) and (\ref{Hunfree2-UG}) of
$\buildrel\ast\over{g}_{ij}$ and $N^i$ define the transformations of $4d$
unimodular metrics  $g_{\mu\nu}$ as it is
parametrized
by the ADM
variables (\ref{ADM}) and (\ref{Lapse}). The transformations
(\ref{Hunfree1g-UG}) and (\ref{Hunfree2-UG}) involve, besides
$\buildrel\ast\over{g}_{ij}$ and $N^i$, also the canonical momenta
$\Pi^{ij}$. The momenta are defined by the Hamiltonian equations as
functions of the ADM variables and the time derivatives of $3d$
metrics:
\begin{eqnarray}
&\partial_0\!\buildrel\ast\over{g}_{ij}
=
\left\{\buildrel\ast\over{g}_{ij}\ ,\
\mbox{$\int$} d^3x\, \mathcal{H}_T\right\}
\Rightarrow&
\nonumber
\\
&\label{Pi}
\Pi^{ij}
=
-\frac{1}{2}
\buildrel\ast\over{g}
\mathcal{G}^{ij\, kl}
\left( \partial_0\!\buildrel\ast\over{g}_{kl}
-2\,\partial_k N_l
+2\buildrel\ast\over{\Gamma}{\!}_{kl}^s N_s\right)\!.&
\end{eqnarray}
As a result, upon the change of parameters
$\epsilon,\epsilon^i\mapsto \xi^\mu$ [see (\ref{xi-epsislon})], the
Hamiltonian gauge transformations (\ref{Hunfree1g-UG}) and
(\ref{Hunfree2-UG}) define the diffeomorphism of $4d$ unimodular
metrics:
\begin{equation}\label{Diff}
\delta_\epsilon g_{\mu\nu} (\buildrel\ast\over{g}_{ij},N^i)
\bigg\vert_{\displaystyle
\begin{array}{l}
\epsilon\mapsto\epsilon(\xi)
\\
\Pi\mapsto\Pi(\buildrel\ast\over{g}_{ij},N^i,\partial_0\!\buildrel\ast\over{g}_{ij})
\end{array}}
=-\nabla_\mu\xi_\nu-\nabla_\nu\xi_\mu  .
\end{equation}
So, one can see that the Hamiltonian gauge transformations
(\ref{Hunfree1g-UG})--(\ref{HGPE-UG}) represent indeed the $(1+3)$-split
form of $4d$ volume-preserving diffeomorphism (\ref{Tdiff}).

Now, let us deduce the Hamiltonian form of the reducible gauge
symmetry (\ref{RedTdiff}) of the UG. The general scheme of replacing
the unfree gauge symmetry by the alternative reducible counterpart
with unrestricted gauge parameters is described in the end of
previous section. It begins with absorbing the structure coefficient
$V^a_\alpha$ of the involution relations (\ref{irinv4}) by the
secondary constraint [see (\ref{tildetau})]. For the UG, relations
(\ref{irinv4}) read as (\ref{UGTalpha}) and (\ref{UGtau}) with
$V^a_\alpha$ identified as $-\partial_i$. So, the reducible secondary
constraints for the UG are defined as
\begin{equation}\label{tilde-tau-UG}
\widetilde{\tau}_i=-\,\partial_i \tau\,,
\end{equation}
where $\tau$ is given by (\ref{UGtau}). Obviously, conservation of
the primary constraints (\ref{TalphaUG}) identifies
$\widetilde{\tau}_i$ as the secondary ones
\begin{equation}
\left\{ T_i\ ,\ \mbox{$\int$} d^3x\, \mathcal{H}_T \right\}
=  T_j  \partial_i N^j +  \partial_j (T_i N^j) +
\widetilde{\tau}_i\, .
\end{equation}
These constraints conserve
\begin{eqnarray}
\left\{\widetilde{\tau}_i\ ,\
\mbox{$\int$} d^3x\, \mathcal{H}_T\right\}
&=&
\partial_i\partial_j\!\left((-\buildrel\ast\over{g})^{-1}\!
\buildrel\ast\over{g}{}^{jk} T_k\right)
\nonumber
\\
&&\mbox{}+\partial_i (\widetilde{\tau}_j N^j)\,.
\end{eqnarray}

The involution relations of all the constraints can be
represented  in terms of functionals
\begin{eqnarray}
T(\zeta)&=&\int  d^3x \, T_i(x)\,\zeta^i(x)\,,
\nonumber
\\
\label{T-funct-red}
\quad \widetilde{\tau}(\rho')&=&\int  d^3x \, \widetilde{\tau}_i(x)\,\rho'{}^i(x)\,,
\end{eqnarray}
where $\rho'^i(x)$ and $\sigma'^i(x)$ are the test functions being arbitrary
vector densities with weights 1
and $\xi^i(x)$ and $\zeta^i(x)$ are test vector fields. For these functionals, the
involution relations  read
\begin{eqnarray}\label{T-funct-red-inv1}
\{ T(\xi)\ ,\ T(\zeta) \}&=&T\big([\xi , \zeta]\big)\,,\\
\{T(\xi)\ ,\ \widetilde{\tau}(\rho') \}
&=&  \widetilde{\tau}\big([\xi , \rho']\big)\,,\\
\{\widetilde{\tau} (\rho')\ ,\ \widetilde{\tau} ( \sigma') \}&=&
 T\big([\rho',\sigma']\big)\,,
\label{T-funct-red-inv3}
\end{eqnarray}
where the brackets denote the following bilinear skew-symmetric
forms of the test functions:
\begin{eqnarray}
&&
[\xi , \zeta]^i =\xi^j\,\partial_j\zeta^i
-\partial_j \xi^i\, \zeta^j\,,\\
&&
[\xi ,\rho']^i = \xi^i\partial_j\rho'^j\,,
\quad
[\rho', \xi  ]^i = -\xi^i\partial_j\rho'^j\,,\\
&&
[\rho',\sigma']^i =
\frac{1}{\buildrel\ast\over{g}}\, \buildrel\ast\over{g}{\!}^{ij}\,
(\partial_k\rho'^k\, \partial_j  \partial_l\sigma'^l
 - \partial_j  \partial_k\rho'^k \, \partial_l\sigma'^l )\,.
\end{eqnarray}
Here $[\xi , \zeta]^i$ is the vector field, $[\xi , \rho'], [\rho',
\xi ] $ is the vector density of the weight 1, and $[\rho',\sigma']^i$
is the vector\,. The involution relations of the constraints and Hamiltonian read
\begin{eqnarray}
\left\{ T_i\ ,\ \mbox{\normalsize{$\int$}} d^3x\,\mathcal{H}_0\right\} &=&
\widetilde{\tau}_i\,,
\\
\left\{\widetilde{\tau}_i\ ,\ \mbox{\normalsize{$\int$}} d^3x\,\mathcal{H}_0\right\} &=&  \partial_i \partial_j\!
\left(
(-\buildrel\ast\over{g})^{-1}\!
\buildrel\ast\over{g}{}^{jk} T_k\right)\,.
\end{eqnarray}

The secondary constraints (\ref{tilde-tau-UG}) are reducible,
\begin{equation}\label{simpl-Z-Z1-UG}
\varepsilon^{Aij}\, \partial_i\widetilde{\tau}_j\equiv 0 \, ,
\end{equation}
where $\varepsilon^{Aij}$ is totaly antisymmetric, taking values
$\{0, 1, -1\}$.

The role of the null vector $Z_1{}_A^\alpha$  [see (\ref{Z})] is
played by  the  dualized De Rham differential. This null vector
is reducible again. The role of the null vector $Z_2{}_{A_1}^A$
[see (\ref{Z})] is played by $\partial_A$. This reducibility of the UG
Hamiltonian constraints corresponds to the general pattern described
in Sec. 2 [cf. (\ref{Z})].

Given the explicit form of irreducible primary constraints,
reducible secondary ones, and the sequence of
null vectors, the reducible gauge symmetry is constructed for the UG
by the general recipe (\ref{RedGS1})--(\ref{SofS2}):
\begin{eqnarray}
\delta_\varepsilon{\!}\buildrel\ast\over{g}_{ij}
&=&
(\dot{\varepsilon}{}^k
- N^k \partial_s\varepsilon^s
-\varepsilon^{ksA}\,\partial_s \varepsilon_A)
\partial_k{\!}\buildrel\ast\over{g}_{ij} 
\nonumber
\\
&&\mbox{}+\buildrel\ast\over{g}_{ik}{\!}
\partial_j(\dot{\varepsilon}{}^k
- N^k \partial_s\varepsilon^s
-\varepsilon^{ksA}\,\partial_s \varepsilon_A)
\nonumber
\\
&&\mbox{}+\buildrel\ast\over{g}_{jk}{\!}
\partial_i(\dot{\varepsilon}{}^k
- N^k \partial_s\varepsilon^s
-\varepsilon^{ksA}\,\partial_s \varepsilon_A)
\nonumber
\\
\label{54}
&&\mbox{}+\frac{2}{\buildrel\ast\over{g}}\, \mathcal{G}_{ij\, kl}\,
\Pi^{kl}\,
\partial_s\varepsilon^s ,
\end{eqnarray}
\begin{eqnarray}\nonumber
\delta_\varepsilon \Pi^{ij}
&=&
\partial_k \big(\Pi^{ij}
(\dot{\varepsilon}{}^k
- N^k \partial_s\varepsilon^s
-\varepsilon^{ksA}\,\partial_s \varepsilon_A)\big)
\nonumber
\\
&&\mbox{}-\,\Pi^{ik}\partial_k
(\dot{\varepsilon}{}^j
- N^j \partial_s\varepsilon^s
-\varepsilon^{jsA}\,\partial_s \varepsilon_A)
\nonumber
\\
&&\mbox{}-\Pi^{jk}\partial_k
(\dot{\varepsilon}{}^i
- N^i \partial_s\varepsilon^s
-\varepsilon^{isA}\,\partial_s \varepsilon_A)
\nonumber
\\
&&\mbox{}-\bigg(\,
\frac{2}{\buildrel\ast\over{g}}\,
\buildrel\ast\over{g}{\!}_{kl} \Pi^{ik}\Pi^{jl}
-\frac{1}{\buildrel\ast\over{g}}\,  \Pi\, \Pi^{ij}
\nonumber
\\
&&\mbox{}\phantom{@@@}
-\frac{1}{\buildrel\ast\over{g}}\buildrel\ast\over{g}{\!}^{ij}
\,\mathcal{G}_{kl\,nm}\Pi^{kl}\Pi^{nm}
+\buildrel\ast\over{R}{\!}^{ij} \bigg)\,\partial_s\varepsilon^s
\nonumber
\\
\label{55}
&&\mbox{}+\mathcal{G}^{ij\, kl}\, \sqrt{-\buildrel\ast\over{g}} \,
\left(\partial_k\partial_l -
\buildrel\ast\over{\Gamma}{\!}^m_{kl}\,\partial_m
\right) \frac{\partial_s\varepsilon^s}{\sqrt{-\buildrel\ast\over{g}}}\,,
\ \
\end{eqnarray}
\begin{eqnarray}
\delta_\varepsilon N^i
&=&
\frac{d}{dt}(\dot{\varepsilon}{}^i
- N^i \partial_s\varepsilon^s
-\varepsilon^{isA}\,\partial_s \varepsilon_A)
\nonumber
\\
&&\mbox{}+(\dot{\varepsilon}{}^j
- N^j \partial_s\varepsilon^s
-\varepsilon^{jsA}\,\partial_s \varepsilon_A)
\partial_j N^i
\nonumber
\\
&&\mbox{}-N^j\partial_j
(\dot{\varepsilon}{}^i
- N^i \partial_s\varepsilon^s
-\varepsilon^{isA}\,\partial_s \varepsilon_A)
\nonumber
\\
\label{56}
&&\mbox{}+\frac{1}{\buildrel\ast\over{g}}
\buildrel\ast\over{g}{\!}^{ij}
\partial_j\partial_s\varepsilon^s ,
\end{eqnarray}
\begin{equation}\label{57}
\delta_\omega \varepsilon^i =
\varepsilon^{ij A}\partial_j\,\omega_A\, ,
\qquad
\delta_\omega \varepsilon_A = \dot{\omega}_A-\partial_A{\omega}\,,
\end{equation}
\begin{equation}\label{58}
\delta_\eta \omega_A =
\partial_A\eta \, ,
\qquad
\delta_\eta \omega = \dot{\eta}\,.
\end{equation}
By construction, the reducible gauge variation of the ADM variables
(\ref{54})--(\ref{56}) with unrestricted gauge
parameters $\varepsilon^i$ and $\varepsilon_A$ should leave the action
(\ref{SH-LUG}) invariant modulo integral of a total divergence. This
fact can be verified by explicitly varying the action:
\begin{eqnarray}
\delta_\varepsilon\displaystyle S
&\equiv &
\int d^4x\,
\bigg(
\delta_\varepsilon\Pi^{ij}\partial_0{\!}\buildrel\ast\over{g}{\!}_{ij}-
\partial_0\Pi^{ij}\delta_\varepsilon{\!}\buildrel\ast\over{g}{\!}_{ij}
-\delta_\varepsilon \mathcal{H}_T \bigg) 
\nonumber
\\
\label{deltaSUGred}
&\equiv &
\mbox{}- \int d^4x\,
\varepsilon_A\, \varepsilon^{Aij}\partial_j\widetilde{\tau}_i
  \quad(mod \, div).
\end{eqnarray}
Given reducibility of the secondary constraints
(\ref{simpl-Z-Z1-UG}), the action is invariant indeed.

Once the reducible gauge symmetry with the second order derivatives
of unrestricted gauge parameters have been derived for the UG
Hamiltonian action, the question appears about connection of these
transformations with $4d$-covariant reducible symmetry
(\ref{RedTdiff}), (\ref{Gauge-for-Gauge}). To answer this question,
we first establish correspondence between the gauge parameters of the
Hamiltonian form of gauge symmetry $\varepsilon^i, \varepsilon_A,
\omega_A, \omega, \eta$
[see (\ref{54}), (\ref{56})--(\ref{58})]
and their $4d$ counterparts
[see (\ref{RedTdiff}),(\ref{Gauge-for-Gauge})]:
\begin{equation}\label{varepsilon-W}
\varepsilon^i = \frac{1}{2}\,
\varepsilon^{ijk}W_{jk}\,,\quad
\varepsilon_A = W_{0A}\,,
\end{equation}
\begin{equation}
\omega_A = \varphi_A\,,
\qquad
{\omega} = \varphi_{0}\,,
\qquad
\eta=\psi\,.
\end{equation}
Second, the Hamiltonian gauge transformations of the ADM variables
involve the canonical momenta $\Pi^{ij}$, while the $4d$ covariant
transformations do not. Replacing $\Pi^{ij}$ by their on-shell
expressions (\ref{Pi}) in the reducible Hamiltonian gauge
transformations, we recover explicitly covariant transformations
(\ref{RedTdiff}), (\ref{Gauge-for-Gauge}) upon the above
identification of gauge parameters.

\section{Hamiltonian BFV-BRST formalism of the UG}
As we have demonstrated in the previous section, the constrained
Hamiltonian formalism of UG admits two alternative formulations for
gauge symmetry. The first one corresponds to the diffeomorphisms
generated by transverse vector fields (\ref{Tdiff}), while another
one corresponds to the reducible gauge transformations
(\ref{RedTdiff}) with the second order time derivatives of
unrestricted gauge parameters being components of antisymmetric
tensor.  Existence of the alternative parametrizations of this type
is a property of any system with unfree gauge symmetry
\cite{Abakumova2021KL}. Therefore  one can  associate two different
BRST complexes with the same action functional once it enjoys the
unfree gauge symmetry. Given the Hamiltonian and constraints, these
two complexes can be constructed along the usual lines of the
BFV-BRST formalism in the \emph{minimal sector} \cite{Batalin:1983pz},
while the specifics of
unfree gauge symmetry reveals itself in the nonminimal sector
\cite{Abakumova2021KL} involved in the gauge fixing. For GR, the
Hamiltonian BRST invariant action has been first constructed by
Fradkin and Vilkovisky in the unpublished preprint
\cite{Fradkin:1977hw}. The Fradkin-Vilkovisky action included the
four ghost vertex in the gauge fixing terms due to the
field dependence of the structure functions of involution relations.
This work includes the BRST transformations, though  it does not
explicitly present the BRST charge for GR. The latter has been
first explicitly presented in the article \cite{Ashtekar:1987hj}.
For the UG, the Hamiltonian BRST formalism has been
unknown for the irreducible explicitly $\Lambda$-dependent set of
constraints (\ref{TalphaUG}) and (\ref{UGtau}), and it is not known for the
reducible $\Lambda$-independent generating set of constraints
(\ref{TalphaUG}) and (\ref{tilde-tau-UG}).
In this section, we
construct the Hamiltonian BFV-BRST formalism for the UG in both
pictures (the first one with irreducible constraints generating the
transverse diffeomorphisms, and the second one for the reducible
constraints generating the second order reducible gauge symmetry)
proceeding along the lines of the general method recently proposed
in Refs. \cite{Abakumova:2019uoo,Abakumova2021KL}.

Let us begin with introducing the canonical set of ghosts for the
irreducible set of constraints (\ref{TalphaUG}) and (\ref{UGtau}). The
canonical ghost pairs of the minimal sector are assigned to all the
constraints
\begin{eqnarray}
&&\displaystyle \{C^i,\overline{P}_j\}=\delta^i_j\,, \quad \text{gh}\,C^i=-\,\text{gh}\,\overline{P}_i=1\,,
\nonumber
\\
&&\varepsilon(C^i)=\varepsilon(\overline{P}_i)=1\,;
\\
&&\displaystyle \{{C},\overline{{P}}\}=1\,, \quad \text{gh}\,{C}=-\,\text{gh}\,\overline{{P}}=1\,,
\nonumber
\\
&&\varepsilon({C})=\varepsilon(\overline{{P}})=1\,.
\end{eqnarray}
The original Hamiltonian action (\ref{SH-LUG}) involves the primary
constraints $T_i$ and corresponding Lagrange multipliers $N^i$. It
does not explicitly involve the secondary constraint $\tau$, nor
does it contain any independent Lagrange multiplier for $\tau$. This
action enjoys the volume-preserving diffeomorphism
(\ref{Hunfree1g-UG})--(\ref{HGPE-UG}).

Let us discuss now the gauge fixing for the UG in Hamiltonian BRST
formalism. Once the four gauge parameters are constrained by one
equation, to fix this gauge symmetry one has to impose three
independent gauge conditions.\footnote{Imposing three independent
conditions we break explicit Lorentz invariance. To preserve the
Lorentz symmetry explicitly one could consider four redundant gauge
conditions. For the unfree gauge symmetry corresponding procedure of
inclusion the redundant gauges are presented in Ref.
\cite{Kaparulin2019}. In this article the general scheme is
exemplified by the de Donder--Fock condition which is redundant for
the linearized UG. For the nonlinear UG, the de Donder--Fock
conditions are independent, so this gauge cannot be imposed. At the
moment, no redundant Lorentz covariant gauge is known for the full
nonlinear UG.} We are going to project out the three independent
conditions from the Lorentz-covariant gauge of GR.  We start with
generalized de Donder--Fock conditions  of GR \cite{Plebanski1961R}
\begin{equation}\label{FDgauge}
D^\mu =   \partial_\nu \, ( (- g)^\Delta\, g^{\mu\nu }) = 0\,, \quad
\Delta\in\mathbb{R}.
\end{equation}
The Hamiltonian BFV-BRST formalism implies to impose the gauge
conditions in the canonical form which is explicitly resolved
with respect  to
the time derivatives of Lagrange multipliers. Let us express the
gauge conditions (\ref{FDgauge}) in terms of ADM variables (the
special case $\Delta=1$ is skipped) and resolve with respect  to
$\partial_0N,\,\partial_0N^i$:
\begin{widetext}
\begin{eqnarray}\label{D0GR}
\frac{D^0}{2(\Delta-1)N^{2\Delta-3}(-\buildrel\ast\over{g})^\Delta}
&=&
\partial_0 N - N^j\partial_j N 
-\frac{\Delta}{2(\Delta-1)}\,N^2(-\buildrel\ast\over{g})^{-\frac{1}{2}}\,\Pi
+\frac{2\Delta - 1}{2(\Delta-1)}\,N\partial_jN^j
 =0\,,
\\
\label{DiGR}
-\,\frac{D^i + N^iD^0}{N^{2\Delta-2}\, (-\buildrel\ast\over{g})^{\Delta}}
&=&
\partial_0 N^i
-N^2\partial_j\!\buildrel\ast\over{g}{\!}^{j i }
-N^j\partial_jN^i 
-\Delta
(-\buildrel\ast\over{g})^{-1}
\buildrel\ast\over{g}{\!}^{ij}
\bigg(
2 N\partial_j N(-\buildrel\ast\over{g})
+ N^2\partial_j(-\buildrel\ast\over{g})
\bigg) = 0\,.
\end{eqnarray}
\end{widetext}
For the UG, these conditions are simplified upon the account of the
unimodularity condition $(-g)=N^2(-\buildrel\ast\over{g})\\ =1$:
\begin{eqnarray}
\frac{D^0}{2(\Delta-1)(-\buildrel\ast\over{g})^{\frac{3}{2}}}
& = &
\frac{1}{2(\Delta-1)(-\buildrel\ast\over{g})^{\frac{3}{2}}}
\nonumber
\\
&&\mbox{}\times
\bigg(-\Pi + (-\buildrel\ast\over{g})\,\partial_jN^j\bigg)
 =0,\
\label{D0UG}
\\
-\frac{D^i + N^iD^0}{(-\buildrel\ast\over{g})}
& = &
\partial_0 N^i
-(-\buildrel\ast\over{g})^{-1}\partial_j\!\buildrel\ast\over{g}{\!}^{j i }
\nonumber
\\
\label{DiUG}
&&\mbox{}
-N^j\partial_jN^i =0\,.
\end{eqnarray}
One can see that the spatial projection  (\ref{DiUG}) of the de
Donder--Fock Lorentz invariant conditions (\ref{FDgauge}) is an
admissible gauge, while the time component (\ref{D0UG}) does not
involve the time derivative of any Lagrange multiplier in the case of
UG. So, we choose the independent gauge conditions
\begin{equation}\label{gauge-cond}
\partial_0 N^i - \chi^i = 0,
\quad
\chi^i = (- \buildrel\ast\over{g})^{-1}
\partial_j\!\buildrel\ast\over{g}{\!}^{j i }
+ N^j \partial_jN^i.
\end{equation}
This gauge does not depend on momenta, unlike the complete de
Donder--Fock conditions (\ref{D0GR}) in the GR.
 Gauge conditions (\ref{gauge-cond}) imply to introduce
Lagrange multipliers $\pi_i$ being canonically conjugate to $N^i$,
\begin{eqnarray}
&&\displaystyle \{N^i,\pi_j\}=\delta^i_j\,, \quad
\text{gh}\,N^i=-\,\text{gh}\,\pi_i=0\,,
\nonumber
\\
\label{irnonminlambda}
&& \varepsilon(N^i)=\varepsilon(\pi_i)=0\,.
\end{eqnarray}
Given the canonical pairs of the above Lagrange multipliers, we
introduce corresponding ghosts of the nonminimal sector,
\begin{eqnarray}
&&\displaystyle \{P^i,\overline{C}_j\}=\delta^i_j\,, \quad \text{gh}\,P^i=-\,\text{gh}\,\overline{C}_i=1\,,
\nonumber
\\
\label{irnonminghosts}
&&\varepsilon(P^i)=\varepsilon(\overline{C}_i)=1\,.
\end{eqnarray}
Since no  gauge condition is imposed being paired with the
``super-Hamiltonian'' constraint $\tau$ (\ref{UGtau}),
then the corresponding Lagrange multipliers and nonminimal sector ghosts are not introduced.
Given the involution relations, Lagrange multipliers and
ghosts, the complete BRST charge reads
\begin{eqnarray}
Q=\int d^3x \bigg(
&&
T_i\, C^i
+\tau\, C
- \overline{P}_i\,C^j\,\partial_j C^i
-\overline{P}\,\partial_i (C^i C)
\nonumber
\\
\label{Qunfree}
&&
\mbox{}+\overline{P}_i\,(-\buildrel\ast\over{g})^{-1}
\buildrel\ast\over{g}{\!}^{ij}\,C\partial_j C
+\pi_i\, P^i\bigg) .
\end{eqnarray}
Notice that the cosmological constant $\Lambda$ is explicitly
involved in this charge through the secondary constraint $\tau$
defined by relation (\ref{UGtau}). This means, the class of
admissible fields is assumed restricted by the boundary conditions
consistent with particular $\Lambda$.

Let us discuss the BRST invariant extension of the Hamiltonian. In
the minimal ghost sector, given the involution relations
(\ref{49}) and (\ref{50}),
the BRST invariant Hamiltonian of
the UG reads
\begin{equation}\label{HscriptUG}
\mathcal{H} = \int d^3x \left(
\mathcal{H}_0(\phi) -\overline{P}\partial_i
C^i -\overline{P}_i\, (-\buildrel\ast\over{g})^{-1}
\buildrel\ast\over{g}{\!}^{ij}\,\partial_j C\right),
\end{equation}
where $\mathcal{H}_0(\phi)$ is the original Hamiltonian (\ref{HUG}).
This Hamiltonian is BRST-exact modulo the additive constant
$\Lambda$ included in the secondary constraint (\ref{UGtau}),
\begin{equation}\label{HUG-exact}
\mathcal{H}=\int d^3x\, \Lambda
+\{ \int d^3x\,\overline{P}\,,\,Q\} .
\end{equation}
At the level of BRST formalism, this  corresponds to the earlier
noticed fact that the Hamiltonian of UG (\ref{HUG}) reduces on shell
[given the secondary constraint (\ref{UGtau})] to the constant
defined by the field value at single point, or at asymptotics,
rather than by the entire Cauchy surface. Since this complex treats
$\Lambda$ as the predefined parameter, it does not have the BRST
cohomology element corresponding to energy.\footnote{This contrasts
to the alternative BRST complex associated with the reducible form
of the volume preserving diffeomorphisms. This issue is considered
in the final part of this section.}

Now, let us construct the complete gauge fixed BRST-invariant
Hamiltonian. Introduce the gauge fermion which includes independent
gauges (\ref{gauge-cond})
\begin{equation}\label{irredPsi}
\displaystyle \Psi=\int d^3x \left(\overline{C}_i\chi^i
+\overline{P}_i N^i\right) .
\end{equation}
Given the gauge fermion $\Psi$  and the BRST-invariant Hamiltonian
of the minimal sector (\ref{HscriptUG}), the complete gauge-fixed
BRST-invariant Hamiltonian reads
\begin{widetext}
\begin{eqnarray}
H_\Psi &=& \mathcal{H}  + \{ Q , \Psi\}
\nonumber
\\
&=&
\int d^3x\,
\bigg\{\,\mathcal{H}_0(\phi) +T_i\,N^i +\pi_i\,\chi^i
\nonumber
\\
&&\phantom{@@@@@}
\mbox{}
-\overline{P}\partial_i C^i
-\overline{P}_i(-\buildrel\ast\over{g})^{-1}\!
\buildrel\ast\over{g}{\!}^{ij}\,\partial_j C \,
+\overline{P}\,  \partial_i (C\,  N^i)
+\overline{P}_i\,(\partial_j C^i\, N^j
-C^j\,  \partial_jN^i)
\nonumber
\\
&&\phantom{@@@@@}
\mbox{}
-\,\overline{C}_i\, (-\buildrel\ast\over{g})^{-1}
\left(
2\,\partial_j \! \buildrel\ast\over{g}{\!}^{ij}
\buildrel\ast\over{\nabla}{\!\!}_k C^k
+\partial_j \big(\!\buildrel\ast\over{\nabla}{\!}^jC^i
+\buildrel\ast\over{\nabla}{\!}^iC^j\big)
\right)
\nonumber
\\
&&\phantom{@@@@@}
\mbox{}
+\,\overline{C}_i\, (-\buildrel\ast\over{g})^{-1}
\left(
 \partial_j \! \buildrel\ast\over{g}{\!}^{ij}
(-\buildrel\ast\over{g})^{-1}\Pi C
-2\, \partial_j \big((-\buildrel\ast\over{g})^{-1}
\Pi^{ij}C\big)
+ \partial_j\big(\buildrel\ast\over{g}{\!}^{ij}
(-\buildrel\ast\over{g})^{-1}\Pi C\big)
\right)
\nonumber
\\
&&\phantom{@@@@@}
\mbox{}
+\,\overline{C}_i\,(\partial_j N^i \,P^j
+  N^j\,\partial_j  P^i)
+ \overline{P}_i P^i \bigg\}\,.
\end{eqnarray}
\end{widetext}
Let us notice the two distinctions of the UG Hamiltonian $H_\Psi$
from the long known GR counterpart \cite{Fradkin:1977hw}. First, the
UG Hamiltonian includes the term with the original Hamiltonian
$\mathcal{H}_0(\phi)$ (\ref{HUG}) while in the GR the analogous term
is absorbed by the super-Hamiltonian constraint multiplied by
the lapse function $N$ being an independent variable. In the UG, we
do not have this variable, while the super-Hamiltonian term can
be reduced to the constant $\Lambda$ by shifting the gauge Fermion
$\Psi\mapsto\Psi-\int d^3x \overline{P}$ [cf. (\ref{HUG-exact})]. In
the GR, the $\Pi$-squared contribution  cannot be eliminated from
the Hamiltonian $H_\Psi$ by any admissible choice of the gauge
fermion. The second distinction concerns the gauge fixing terms and
related ghosts. In the UG, the relativistic gauge
(\ref{gauge-cond}), being the spatial projection of the Lorentz
invariant condition (\ref{FDgauge}), does not result in the
four ghost vertices unlike the GR counterpart \cite{Fradkin:1977hw}.

To summarize this version of the Hamiltonian BFV-BRST formulation of
the UG, we see that it corresponds to the  fixed asymptotics of the
fields and explicitly involves the corresponding cosmological constant
as a predefined parameter.
It differs, however, from the BFV
formulation of GR by the content of the phase space, and by the
structure of the BRST-invariant gauge-fixed Hamiltonian.
This
version of the BFV-BRST formalism suits well to the interpretation
of the UG as the system with the fixed field
asymptotics such that
it corresponds to a
certain predefined value of the cosmological constant.
If the gravity could be quantized beyond the formal level (our consideration is formal) proceeding from this version of the BRST formalism, the cosmological constant would be involved in the quantum theory just as a numerical parameter whose value is fixed from the outset. This leaves no room for quantum transitions between the states with different values of $\Lambda$.

Let us construct now an alternative BFV-BRST formalism which
proceeds from the reducible set of the secondary constraints
(\ref{tilde-tau-UG}). As we shall see, this formalism does not
assume to fix the field asymptotics, nor does it explicitly involve
the cosmological constant as the predefined parameter.

In the construction of the formalism, we follow the general scheme
of the article \cite{Abakumova2021KL} concerning the BFV-BRST
formalism with reducible \emph{secondary} constraints. This scheme
differs from the classical BFV construction
\cite{Batalin:1983pz} by the nonminimal sector, including ghosts,
Lagrange multipliers, and gauge conditions.

Given the irreducible primary constraints (\ref{TalphaUG}), and
reducible secondary ones (\ref{tilde-tau-UG}) and
(\ref{simpl-Z-Z1-UG}), we introduce the ghosts of the minimal sector,
\begin{eqnarray}
&&\displaystyle \{C^i,\overline{P}_j\}=\delta^i_j\,, \quad \text{gh}\,C^i=-\,\text{gh}\,\overline{P}_i=1\,,
\nonumber
\\
\label{redghosts1}
&&\varepsilon(C^i)=\varepsilon(\overline{P}_i)=1\,;
\end{eqnarray}
\begin{eqnarray}
&&\displaystyle \{\mathcal{C}^i,\overline{\mathcal{P}}_j\}=\delta^i_j\,, \quad \text{gh}\,\mathcal{C}^i=-\,\text{gh}\,\overline{\mathcal{P}}_i=1\,,
\nonumber
\\
\label{redghosts2}
&&\varepsilon(\mathcal{C}^i)=\varepsilon(\overline{\mathcal{P}}_i)=1\,;
\end{eqnarray}
\begin{eqnarray}
&&\{\mathcal{C}{}_A,\overline{\mathcal{P}}{}^{B}\}=\delta^B_A\,, \quad \text{gh}\,\mathcal{C}{}_A=-\,\text{gh}\,\overline{\mathcal{P}}{}^{A}=2\,,
\nonumber
\\
\label{gg1}
&&\varepsilon(\mathcal{C}{}_A)=\varepsilon(\overline{\mathcal{P}}{}^{A})=0\,;
\end{eqnarray}
\begin{eqnarray}
&&\{\mathcal{C},\overline{\mathcal{P}}\}=1\,, \quad \text{gh}\,\mathcal{C}{}=-\,\text{gh}\,\overline{\mathcal{P}}=3\,,
\nonumber
\\
\label{gg2}
&&\varepsilon(\mathcal{C})=\varepsilon(\overline{\mathcal{P}})=1\,.
\end{eqnarray}

The shift functions $N^i$, being the Lagrange multipliers to the \emph{primary} constraints,  are complemented by the conjugate momenta
\begin{eqnarray}
&&\{N^i,\pi_j\}=\delta^i_j\,, \quad \text{gh}\,N^i=-\,\text{gh}\,\pi_i=0\,,
\nonumber
\\
\label{N-momenta}
&&\varepsilon(N^i)=\varepsilon(\pi_i)=0\,.
\end{eqnarray}
These momenta are to serve as Lagrange multipliers to the \emph{irreducible} gauge conditions (\ref{gauge-cond}).

The canonical conjugate pairs of ghosts are introduced for the irreducible gauge conditions
\begin{eqnarray}
&&\{P^i,\overline{C}_j\}=\delta^i_j\,, \quad \text{gh}\,P^i=-\,\text{gh}\,\overline{C}_i=1\,,
\nonumber
\\
\label{PrimaryGhosts-nonmin-reduce}
&&\varepsilon(P^i)=\varepsilon(\overline{C}_i)=1\,.
\end{eqnarray}
As far as the primary constraints are concerned, the nonminimal sector is constructed following the pattern of the irreducible gauge symmetry and the gauge fixing without redundancy.

Now, we turn to the nonminimal sector related to the reducible secondary constraints (\ref{tilde-tau-UG}). Neither are Lagrange multipliers  present in the theory for these constraints nor are gauge conditions  imposed being paired with these constraints. So, the nonminimal sector of the secondary constraints does not include the canonical pairs of Lagrange multipliers, nor are introduced the ghosts related to the gauge conditions. Once the secondary constraints are  reducible, they generate the redundant gauge symmetry (\ref{54})--(\ref{58}) of the original variables. This leads to the ghosts for ghosts in the minimal sector, and this requires one to impose the gauge conditions on the ghosts related to the secondary constraints and their reducibilities. These gauge conditions, in their own turn, require corresponding extra ghosts. These are introduced following the pattern of redundant gauge conditions in the sector of the gauge of the ghosts, while the Lagrange multipliers and related ghosts are not introduced in this sector.

The nonminimal sector ghosts of the first reducibility of secondary constraints read
\begin{eqnarray}
&&\{\mathcal{P}{}_A,\overline{\mathcal{C}}{}^{B}\}=\delta_A^B\,, \quad \text{gh}\,\mathcal{P}{}_A=-\,\text{gh}\,\overline{\mathcal{C}}{}^{A}=2\,,
\nonumber
\\
\label{extrag1}
&&\varepsilon(\mathcal{P}{}_A)=\varepsilon(\overline{\mathcal{C}}{}^{A})=0\,.
\end{eqnarray}
Also Lagrange multipliers are introduced to the first level reducibility and corresponding gauge conditions imposed on the original ghosts for reducible secondary constraints,
\begin{eqnarray}
&& \{\lambda_A,\pi^B\}=\delta_A^B\,, \quad
\text{gh}\,\lambda_A=-\,\text{gh}\,\pi^A=1\,,
\nonumber
\\
&&\label{etral1}
\varepsilon(\lambda_A)=\varepsilon(\pi^A)=1\,.
\end{eqnarray}
A similar set of the nonminimal sector ghosts and multipliers is introduced at the second reducibility level
\begin{eqnarray}
&&
\{\mathcal{P},\overline{\mathcal{C}}\}=1\,,
\quad
\text{gh}\,\mathcal{P}=-\,\text{gh}\,\overline{\mathcal{C}}=3\,,
\nonumber
\\
&&\label{extrag2}
\varepsilon(\mathcal{P})=\varepsilon(\overline{\mathcal{C}})=1\,;
\end{eqnarray}
\begin{eqnarray}
&&\{\lambda,\pi\}=1\,, \quad
\text{gh}\,\lambda=-\,\text{gh}\,\pi=2\,,
\nonumber
\\
&&\label{extral2}
\varepsilon(\lambda)=\varepsilon(\pi)=0\,.
\end{eqnarray}

The second reducibility also requires one to introduce the extra ghosts
\begin{eqnarray}
&& \{\lambda^{(1')},\pi^{(1')}\}=1\,, \quad \text{gh}\,\lambda^{(1')}=-\,\text{gh}\,\pi^{(1')}=1\,,
\nonumber
\\
&&\varepsilon(\lambda^{(1')})=\varepsilon(\pi^{(1')})=1\,;
\end{eqnarray}
\begin{eqnarray}
&&\{\mathcal{P}^{(1')},\overline{\mathcal{C}}{}^{(1')}\}=1\,, \quad \text{gh}\,\mathcal{P}^{(1')}=-\,\text{gh}\,\overline{\mathcal{C}}{}^{(1')}=2\,,
\nonumber
\\
&&\label{P1'A1C1'B1}
\varepsilon(\mathcal{P}^{(1')})=\varepsilon(\overline{\mathcal{C}}{}^{(1')})=0\,.
\end{eqnarray}

Given the ghost and Lagrange multiplier spectrum, constraints, and related null vectors, the complete BRST charge for the reducible gauge symmetry of UG reads
\begin{eqnarray}
Q
=
\displaystyle\int d^3 x \bigg(
&&
T_i\, C^i
+\widetilde{\tau}_i\, \mathcal{C}^i
+\overline{\mathcal{P}}_i\,\varepsilon^{ij A} \,\partial_j\mathcal{C}_A
+\overline{\mathcal{P}}{}^A\, \partial_A \mathcal{C}
\nonumber
\\
&&
\displaystyle\mbox{}
- \overline{P}_i\,C^j\partial_j\, C^i
-\overline{\mathcal{P}}_i\,
C^i\partial_j\,\mathcal{C}^j
\nonumber
\\
&&
\displaystyle\mbox{}
+\overline{P}_i\,
(-\buildrel\ast\over{g})^{-1}\!
\buildrel\ast\over{g}{\!}^{ij}\,
\partial_k\mathcal{C}^k\,\partial_j\partial_l\mathcal{C}^l
\nonumber
\\
&&
\displaystyle\mbox{}
+\frac{1}{2}\,\overline{\mathcal{P}}{}^A\,
\varepsilon_{Aij}\,C^i C^j\,\partial_k\mathcal{C}^k
\nonumber
\\
&&
\displaystyle\mbox{}
+\frac{1}{6}\,\overline{\mathcal{P}}\, \varepsilon_{ijk}
C^i C^j C^k\,\partial_l\mathcal{C}^l
\nonumber
\\
&&
\mbox{}+\pi_i P^i
+\pi^A\mathcal{P}_A
\nonumber
\\
&&
\mbox{}+\pi\mathcal{P}
+\pi^{(1')}\mathcal{P}^{(1')}\bigg),
\label{Qred}
\end{eqnarray}
where $\varepsilon_{ijk}$ is totally antisymmetric
and takes values $\{0, 1, -1\}$, and
$\varepsilon^{ij k}\varepsilon_{ksm} =
-(\delta^i_s\delta^j_m
- \delta^i_m\delta^j_s)$.

This BRST charge involves the ghost terms up to the fifth order even though the constraint algebra is closed, see involution relations (\ref{T-funct-red-inv1})--(\ref{T-funct-red-inv3}).
These higher order ghost terms are related to the off-shell disclosure of the  algebra of reducible constraints: the null vectors are involved in the compatibility conditions for the structure functions of the involution relations.

In the minimal sector, the BRST invariant extension $\mathcal{H}$ of the original UG's Hamiltonian $\mathcal{H}_0$ reads
\begin{eqnarray}
\mathcal{H}
=
\int d^3 x\bigg(
&&
\mathcal{H}_0(\phi)
-\overline{\mathcal{P}}_i\,C^i
-\overline{P}_i\,(-\buildrel\ast\over{g})^{-1}\!
\buildrel\ast\over{g}{}^{ij}\, \partial_j\partial_k\mathcal{C}^k
\nonumber
\\
&&
\mbox{}
+\frac{1}{2}\ \overline{\mathcal{P}}{}^A \varepsilon_{Aij}C^iC^j
\nonumber
\\
&&
\mbox{}
+\frac{1}{6}\, \overline{\mathcal{P}}\varepsilon_{ijk}C^iC^jC^k\bigg).
\end{eqnarray}
Gauge fixing implies to involve the nonminimal sector through the gauge fermion $\Psi$,
\begin{equation}
H_\Psi = \mathcal{H}+\{Q,\Psi\}\,.
\end{equation}
We suggest to choose the gauge Fermion in the following way:
\begin{eqnarray}
\displaystyle \Psi
=
\int d^3x
\bigg(
&&
\overline{C}_i\, \chi^i
+\overline{P}_i\, N^i
\nonumber
\\
&&
\mbox{}
+(-\buildrel\ast\over{g}{\!})^{-1}\,
\overline{\mathcal{C}}{}^A
\varepsilon_{Aji}\!
\buildrel\ast\over{\nabla}{\!\!}^j\mathcal{C}^i
+\overline{\mathcal{P}}{}^A\,\lambda_A
\nonumber
\\
&&
\mbox{}
+(-\buildrel\ast\over{g}{\!})^{-1}\,
\overline{\mathcal{C}}
\buildrel\ast\over{\nabla}{\!\!}^A \mathcal{C}_A
+\overline{\mathcal{P}}\, \lambda\
\nonumber
\\
&&
\mbox{}
-\, (-\buildrel\ast\over{g}{\!})^{-1}
\buildrel\ast\over{\nabla}{\!\!}^A
\overline{\mathcal{C}}{}^{(1')}
\, \lambda_A
+\overline{\mathcal{C}}{}^A\!
\buildrel\ast\over{\nabla}{\!\!}_A\lambda^{(1')}
\nonumber
\\
&&
\mbox{}
-\overline{\mathcal{C}}
\,\mathcal{P}^{(1')}
\bigg).
\label{Psired}
\end{eqnarray}
This choice breaks the
spatial reparametrization
invariance in the sector of original variables (as the de Donder--Fock conditions $\chi^i$ are involved), while  it involves ghosts and ghost for ghosts in a reparametrization invariant way.

Given the gauge fermion, the gauge-fixed BRST invariant Hamiltonian of UG reads:
\begin{widetext}
\begin{eqnarray}
H_\Psi= \int d^3x\,
\bigg\{
&&
\mathcal{H}_0(\phi) +T_i\,N^i +\pi_i\,\chi^i 
\nonumber
\\
&&
\displaystyle\mbox{}
+\left(\varepsilon^{Aji}
\buildrel\ast\over{\nabla}{\!\!}_j\overline{\mathcal{P}}_i
- (-\buildrel\ast\over{g}{\!})^{-1}
\buildrel\ast\over{\nabla}{\!\!}^A
\pi^{(1')}
\right)\lambda_A\,
-\buildrel\ast\over{\nabla}{\!\!}_A\overline{\mathcal{P}}{}^A
\lambda\, 
\nonumber
\\
&&
\displaystyle\mbox{}
+\pi^A \left(
(-\buildrel\ast\over{g}{\!})^{-1}
\varepsilon_{Aji}
\buildrel\ast\over{\nabla}{\!}^j
\mathcal{C}^i
+\buildrel\ast\over{\nabla}{\!\!}_A\lambda^{(1')}
\right)
+\pi\left(
(-\buildrel\ast\over{g}{\!})^{-1}\!
\buildrel\ast\over{\nabla}{\!\!}^A \mathcal{C}_A
-\mathcal{P}^{(1')}
\right)
\nonumber
\\
&&
\displaystyle\mbox{}
-\overline{\mathcal{P}}_i (C^i-N^i\partial_k\mathcal{C}^k)
-\overline{P}_i\,(-\buildrel\ast\over{g})^{-1}\!
\buildrel\ast\over{g}{}^{ij}\, \partial_j\partial_k\mathcal{C}^k
+\overline{P}_i\,(\partial_j C^i\,N^j-C^j\,\partial_j N^i)\,
\nonumber
\\
&&
\displaystyle\mbox{}
-\,\overline{C}_i\, (-\buildrel\ast\over{g})^{-1}
\left(
\partial_j \! \buildrel\ast\over{g}{\!}^{ij} X
+\partial_j X^{ij}
\right)
+\, \overline{C}_i\,(\partial_j N^i \,P^j+N^j  \partial_j P^i)\,
\nonumber
\\
&&
\displaystyle\mbox{}
+\overline{\mathcal{C}}{}^A
\left(
(-\buildrel\ast\over{g}{\!})^{-1}\!
\big(\delta_A^B\!
\buildrel\ast\over{\Delta}
-\buildrel\ast\over{\nabla}{\!\!}^B\!
\buildrel\ast\over{\nabla}{\!\!}_A
\big)\mathcal{C}_B
\,+\buildrel\ast\over{\nabla}{\!\!}_A\mathcal{P}^{(1')}
\right)
+\overline{\mathcal{C}}\,(-\buildrel\ast\over{g}{\!})^{-1}\!
\buildrel\ast\over{\Delta}\mathcal{C}\, 
\nonumber
\\
&&
\displaystyle\mbox{}
+\overline{P}_i\,P^i
+\left(
\overline{\mathcal{P}}{}^A
-(-\buildrel\ast\over{g}{\!})^{-1}\!
\buildrel\ast\over{\nabla}{\!\!}^A
\overline{\mathcal{C}}{}^{(1')}
\right)
\mathcal{P}_A
+\overline{\mathcal{P}}\,\mathcal{P}\,
\nonumber
\\
&&
\displaystyle\mbox{}
+\frac{1}{2}\,\overline{\mathcal{P}}{}^A\,
\varepsilon_{Aij}\,
(C^iC^j-2\,C^iN^j\partial_k\mathcal{C}^k)
-\,\overline{\mathcal{C}}{}^A
(-\buildrel\ast\over{g}{\!})^{-1}
\varepsilon_{Aji}
\buildrel\ast\over{\nabla}{\!}^j
(C^i\partial_k\mathcal{C}^k)\,
\nonumber
\\
&&
\displaystyle\mbox{}
+ \overline{\mathcal{C}}{}^A
(-\buildrel\ast\over{g}{\!})^{-1}
\varepsilon_{Aji}\!
\left(\!
X\!\buildrel\ast\over{\nabla}{\!}^j
\mathcal{C}^i
+X^{jk}\!\buildrel\ast\over{\nabla}{\!\!}_k\mathcal{C}^i
-\buildrel\ast\over{\nabla}{\!}^j X^i{}_k\,\mathcal{C}^k
+\frac{1}{2}\buildrel\ast\over{\nabla}{\!}^j X\, \mathcal{C}^i\!\right)
-\, \overline{\mathcal{C}}{}^A
\buildrel\ast\over{\nabla}{\!\!}_AX\,\lambda^{(1')}\,
\nonumber
\\
&&
\displaystyle\mbox{}
-\left(
\buildrel\ast\over{\nabla}{\!\!}^A
\overline{\mathcal{C}}{}^{(1')} X
+\buildrel\ast\over{\nabla}{\!\!}_i
\overline{\mathcal{C}}{}^{(1')}
X^{iA}
+\frac{3}{2}\,
\overline{\mathcal{C}}{}^{(1')}\!
\buildrel\ast\over{\nabla}{\!\!}^A\!X
\right)
(-\buildrel\ast\over{g}{\!})^{-1}
\,\lambda_A\,
\nonumber
\\
&&
\displaystyle\mbox{}
-\overline{\mathcal{C}}
(-\buildrel\ast\over{g}{\!})^{-1}
\bigg(
X\buildrel\ast\over{\nabla}{\!\!}^A\mathcal{C}_A
\,+\buildrel\ast\over{\nabla}{\!\!}_i
\big(X^{iA}\,\mathcal{C}_A\big)
-\frac{1}{2}\buildrel\ast\over{\nabla}{\!\!}^A\!X\,\mathcal{C}_A
\bigg)\,
\nonumber
\\
&&
\displaystyle\mbox{}
+\frac{1}{6}\,\overline{\mathcal{P}}\, \varepsilon_{ijk}
(C^i C^jC^k -3\,C^i C^j N^k\partial_l\mathcal{C}^l)
+\frac{1}{2}\,
\overline{\mathcal{C}}\,
(-\buildrel\ast\over{g}{\!})^{-1}
\varepsilon_{Aij}\!
\buildrel\ast\over{\nabla}{\!\!}^A
(C^i C^j\,\partial_k\mathcal{C}^k)\bigg\}\,,
\label{HPsired}
\end{eqnarray}
\end{widetext}
where $\buildrel\ast\over{\Delta}=\buildrel\ast\over{\nabla}{\!\!}_i\!
\buildrel\ast\over{\nabla}{\!}^i$
and the following abbreviation is used
\begin{eqnarray}
X^{ij}
&=&
\buildrel\ast\over{\nabla}{\!}^iC^j
+\buildrel\ast\over{\nabla}{\!}^jC^i
+(-\buildrel\ast\over{g})^{-1}
\left(
2\,\Pi^{ij}- \buildrel\ast\over{g}{\!}^{ij} \Pi
\right)\partial_k\mathcal{C}^k ,
\nonumber
\\
X &=& \buildrel\ast\over{g}_{ij}\!X^{ij}=
2\buildrel\ast\over{\nabla}{\!\!}_k C^k
-(-\buildrel\ast\over{g})^{-1}\Pi\partial_k\mathcal{C}^k .
\label{abrX}
\end{eqnarray}
Covariant derivatives in
(\ref{Psired}), (\ref{HPsired}), and (\ref{abrX})
are defined as for tensor densities of corresponding weight
\begin{eqnarray}
p(\,\buildrel\ast\over{g}_{ij}\,, \Pi^{ij})
&=&
p(C^i, \overline{P}_i)
=p(\mathcal{C}_A , \overline{\mathcal{P}}{}^A)
\nonumber
\\
&=&
p(\mathcal{C} , \overline{\mathcal{P}})
=(0,1),
\nonumber
\\
p(\mathcal{C}^i , \overline{\mathcal{P}}_i)
&=&(1,0),
\nonumber
\\
p(N^i , \pi_i)
&=&
p(P^i, \overline{C}_i)
=p(\lambda_A , \pi^A)
=p(\mathcal{P}_A , \overline{\mathcal{C}}{}^A)
\nonumber
\\
&=&p(\lambda , \pi)
=p(\mathcal{P} , \overline{\mathcal{C}})
=(-1,2),
\nonumber
\\
p(\,\lambda^{(1')}\,,\,\pi^{(1')})
&=&
p(\,\mathcal{P}^{(1')}\,,\,\overline{\mathcal{C}}{}^{(1')})
=(-2,3),
\nonumber
\\[2mm]
p((-\buildrel\ast\over{g}{\!})^{\frac{1}{2}})
&=&
p(\varepsilon^{ijk})=-p(\varepsilon_{ijk})=1.
\end{eqnarray}
As we have seen in this section, the same classical Hamiltonian action of the UG (\ref{SH-LUG}) and (\ref{HUG}) gives rise to two different Hamiltonian BRST formalisms. The first one involves the irreducible secondary constraints (\ref{UGtau}). Corresponding constraint algebra generates the unfree gauge
symmetry with the gauge parameters obeying the transversality condition
(see in Sec. 3). The irreducible secondary constraints (\ref{UGtau}), and hence the BRST charge, explicitly involve the cosmological constant as the parameter predefined by the field asymptoptics. So, the cosmological constant is not a
physical quantity in this setup as no BRST cocycle is associated with $\Lambda$. Another BRST complex, being based on the reducible set of secondary constraints corresponds to the reducible form of the volume preserving diffeomorphisms
(see in Sec. 3).  The reducible constraints (\ref{tilde-tau-UG}) do not explicitly
involve $\Lambda$. Corresponding physical quantity $\tau$ is a cocycle of the local BRST complex generated by the BRST charge (\ref{Qred}). So, this BRST
complex captures the UG dynamics with various cosmological constants.
If the gravity could be quantized beyond the formal level (we do not discuss here the notorious problem of renormalizing quantum gravity) proceeding from this BRST formulation, this would mean that the initial quantum state could be a mixture of various cosmological constants, and the quantum transitions are admissible between the states with different $\Lambda$'s.
The choice between these two inequivalent BRST-complexes depends on the setup of the physical problem. If the asymptotics of the fields is predefined (hence, $\Lambda$ is fixed from the outset) the first option has to be chosen. If various asymptotics are admitted for the metrics, the second option is chosen where the cosmological constant is BRST cocycle, and hence it can enjoy dynamics at quantum level.

\section{Conclusion}
Let us briefly summarize the results and discuss further
perspectives.

For the UG, we have found the Hamiltonian form of the volume preserving
diffeomorphism transformations with the gauge parameters restricted
by the transversality equations.  We also find the Hamiltonian
counterpart of the  reducible form of UG gauge symmetry with the
unrestricted gauge parameters (\ref{RedTdiff}) enjoying gauge
symmetry of their own. Proceeding from these two alternative
Hamiltonian descriptions of the volume preserving diffeomorphisms,
we construct two alternative Hamiltonian BFV-BRST formalisms for
the UG. These constructions are worked out along the lines of the
recent article \cite{Abakumova2021KL} which formulates the
Hamiltonian description for a general system with unfree gauge
symmetry. Let us mention some specifics of these BFV-BRST
formalisms. The first of them, being related to the unfree form of
the gauge symmetry, explicitly involves the cosmological constant
$\Lambda$. This implies to fix the asymptotics of the fields, and
the constant $\Lambda$ is a fixed parameter from the viewpoint of
this complex rather than an element of the cohomology group. Also
it is
interesting to note that this form of the UG Hamiltonian BFV-BRST
formalism, being very close to the GR analog, admits such a
projection of the de Donder--Fock condition that does not result in
the four ghost vertex in the gauge fixed BRST invariant Hamiltonian,
while for the GR this higher-order ghost vertex is inevitable for
any known relativistic gauge fixing. The second form of the
BFV-BRST Hamiltonian formalism corresponds to the reducible form of
the UG gauge symmetry with unrestricted gauge parameters
(\ref{RedTdiff}). In this picture, the BRST charge does not
explicitly involve the cosmological constant, nor does the BRST
invariant Hamiltonian. From the standpoint of the latter complex,
the cosmological constant is a nontrivial element of the BRST
cohomology group. This corresponds to the interpretation of
$\Lambda$ as the ``global degree of freedom'' while the former BRST
complex treats the constant as a fixed parameter. In the
BV formulation of the UG with the reducible
parametrization
of gauge symmetry (\ref{RedTdiff}), the cosmological constant is
treated as the element of the local BRST cohomology group in the
recent work \cite{Elfimov:2021aok}.

Let us mention that any case of unfree gauge symmetry leads to the
``global conserved quantities''
\cite{Abakumova:2020ajc,Abakumova2021KL,Abakumova:2021rlq}.
Equally uniform
for any unfree symmetry is the existence of an alternative form of
gauge transformations with unrestricted gauge parameters enjoying
the gauge symmetry of their own \cite{Abakumova2021KL}. Various
modifications of the UG are discussed in the literature (see
\cite{BARVINSKY201759,Barvinsky-Kolganov,Jirousek:2018ago,Jirousek:2020vhy}
 and references
therein) where the cosmological constant, or even the Newtonian one,
arises as the ``global conserved quantity". If one is going to treat
these constants as the degrees of freedom, not as the parameters
fixed by the predefined asymptotics of the fields, the reducible
form of the gauge symmetry seems preferable.

\begin{acknowledgments}
The  part of this work related to the BRST formalism for UG is supported by the Foundation for the Advancement of Theoretical Physics and Mathematics “BASIS”.  The study of constrained Hamiltonian formalism for the UG is supported by the Tomsk State University Development Program (Priority-2030).
\end{acknowledgments}


\end{document}